# StegHash: New Method for Information Hiding in Open Social Networks

Krzysztof Szczypiorski
Warsaw University of Technology, Warsaw, Poland
Cryptomage SA, Wrocław, Poland
e-mail: ksz@tele.pw.edu.pl

*Abstract*—**In this paper a new method for information hiding in open social networks is introduced. The method, called StegHash, is based on the use of hashtags in various open social networks to connect multimedia files (like images, movies, songs) with embedded hidden messages. The evaluation of the system was performed on two social media services (Twitter and Instagram) with a simple environment as a proof of concept. The experiments proved that the initial idea was correct, thus the proposed system could create a completely new area of threats in social networks.**

## I. INTRODUCTION

Steganography seems to be a very promising technology for sharing information, especially in the time "before" post quantum cryptography, when there is still a need for the design of tools to communicate securely and no certainty that most of the contemporary cryptography will survive. As observed in [1] recently, major attention has been paid to constructing image [2] and network [3] steganography methods. Lately, less effort has been applied to text steganography [4], so this work revisited this attractive area for research in combination with social media.

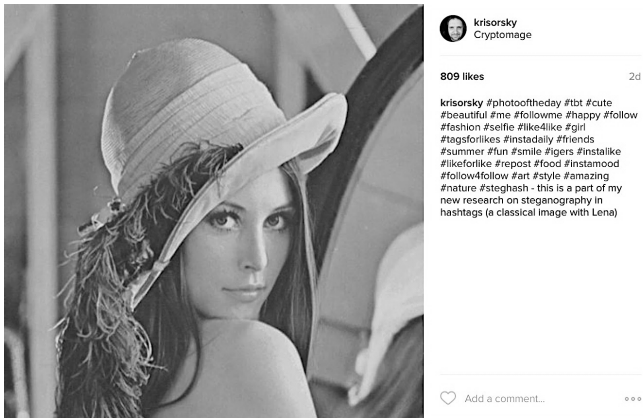

**Figure 1 Usage of hashtags in Instagram – https://www.instagram.com/p/BJVhaADBbT9 .**

In this paper, a new method for hiding information in open social networks (OSNs), called StegHash (Steganographic Hashtags), is introduced. A hashtag is typically a label containing a word starting with the "#" (hash) symbol that is attached to a message posted on social networks. Figure 1 contains a classical image of Lena tagged with 30 hashtags from Instagram. According to [5] "social media is natural platform for the spread of thoughts and ideas, sometimes called memes" and hashtags could be consider as potential memes, especially on platforms with length restrictions for the messages (like Twitter that is 140 letters). Therefore, hashtags are not only limited to regular words from dictionaries, but also could be combinations of acronyms and linguistic circus skills (like #legs2die4, #like4like). With almost no limits for the construction of hashtags, due to thousands of languages worldwide with dozens (or even hundreds) of alphabets, the infinite world of indexes could be explored for more than a lifetime.

In our work we abstract from the linguistic level and forget the exact meaning of the hashtags as understood by humans. The proposed method of StegHash is based on the use of hashtags on various social networks to connect multimedia files, like images, movies, or songs, with embedded hidden messages. For every set of hashtags containing $n$ elements there is the factorial of $n$ permutations, which are individual indexes of each message. Having a secret value (password) and a secret transition generator (function) the link between these indexes could be established and then explored as a chain from one message to another, with each containing hidden content.

To prove that the idea of StegHash is correct, a simple evaluation environment was prepared to inject messages into two popular OSNs (Twitter and Instagram). We choose a hoping technique from one service to the other, just to show how many possibilities come with StegHash. Every service has different features and policies on sanitizing the uploaded content. Therefore, for many reasons it is easier to use image steganography on Google Plus than on Facebook [6][7].

Primarily our motivation for this work was to find new threats or anomalies that could be analyzed and then detected only by big data algorithms, rather than small data ones, and this is why steganography in OSNs was an excellent topic for this purpose. We would like to further our two previous efforts: the first on perfect undetectability [8] and the second on steganographic routing [9]. In [8] we applied the same approach for constructing steganographic algorithms as was used for symmetric encryption ones and proved that it was hard to perform. The work presented in [9] was the first attempt in the literature to use many different carriers (like image, text, movie, audio, network steganography) to bypass existing security systems, but it was designed following military requirements (mobile agent system technology), hence it was too hard for real life applications.

This paper is structured as follows: Section II briefly presents the state of the art in social network steganography,

including a background to text steganography. Section III contains a presentation of the idea of the StegHash method and a typical scenario for the preparation of the steganograms. In Section IV the work describes a proof of concept and shows the initial results. Section VI includes a discussion on the possibility of the detection of the proposed system. Finally, Section VII concludes our efforts and suggests future work.

## II. RELATED WORK

In [10], Beato et al. presented two models of communication: high-entropy and low-entropy. The high-entropy model utilizes media such as pictures, video, and music, etc. to embed steganographic messages. In this model the steganogram is transported by a single object. This is a classic method of steganographic communication, in which a steganogram is applied as part of the picture. In this model, the steganographic throughput is high but the channel is easy to detect. The second model is based on a null cipher approach. It utilizes text data (e.g., status update, group text message) to carry secret information. The mechanism to determine the steganogram location relays on a pre-shared secret to decode the actual message. The suggested appliance is mainly signaling due to the low steganographic throughput. The authors proposed utilizing such a covert channel to determine the actual steganogram location, which can be part of another online service.

Castiglione et al. presented in [11] two low-entropy steganographic methods. The first method utilizes filenames to carry hidden messages and requires an OSN that does not change the filename. The authors proposed utilizing the default naming schemes of popular digital camera producers and a photo sequence number to carry the hidden message. This method has a relatively small steganographic throughput but is hard to detect. The second method takes advantage of the feature of inserting tags in images. The proposed stealth communication channel requires the uploading of multiple images and to tag multiple users. Based on a predefined image and user sequence, a binary matrix can be determined. The second method has a relatively low steganographic throughput.

Wilson et al. [12] and Champan et al. [4] presented linguistic approaches to hide information in twitter posts. Steganograms are carried by a bitmap determined by a language permutation. Such a channel is considered to be very secure, although it requires a human review of tweets and has a very small steganographic throughput.

All proposed methods utilize either a classic image steganography approach, which can be detected easily, or more sophisticated methods, for which the steganographic throughput is relatively small. For example, sending X bytes of data using image user tags requires uploading Y images and tagging Z users. The other disadvantage in the proposed methods is the fact that a steganogram sender is linked with the various user accounts that he/she or someone else are required to open. Such behavior can arise suspicions (OSN providers utilize algorithms that detect when someone tries to open many accounts).

All of the state-of-the-art methods are designed to operate on a single OSN, except the signaling channels presented in [10].

## III. IDEA OF STEGHASH

The proposed method is based on the use of hashtags in the OSNs to connect multimedia files (like images, movies, songs) with embedded hidden messages. The set of hashtags is the base for constructing the indexes, which are unique labels to mark up each update in the OSN. For every set of hashtags containing $n$ elements there is the factorial of $n$ permutations, and every single instance produces an individual index for a given message. Having a secret value (a password) and a secret transition generator, the link between these indexes could be established and then explored as a chain from one message to another with each containing hidden content (Fig. 2). The set of hashtags is independent from the OSN technology and could also be used on regular web pages. The key issue is how to determine the placement in next message? A search engine designed for OSNs should be used, due to its capacity to search the hashtags as a primary way of marking messages in the social media. In addition, the built in search option of the given OSN could be used.

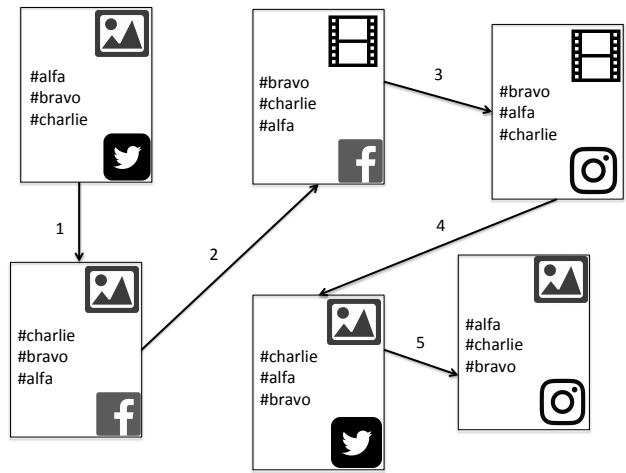

**Figure 2 Example of StegHash method.**

Let $l$ be the length of an address in bits for creating the index for the group containing $n$ hashtags:

$$l = \lceil log_2 n! \rceil, n > 1 \qquad (1)$$

TABLE I. NUMBER OF PERMUTATIONS AND LENGTH OF ADDRESS IN FUNCTION OF N.

| n | n! | l | wasted |
|---|-----|----|--------|
| 2 | 2   | 1  | 0.0%   |
| 3 | 6   | 3  | 33.3%  |
| 4 | 24  | 5  | 33.3%  |
| 5 | 120 | 7  | 6.7%   |
| 6 | 720 | 10 | 42.2%  |

| n | n! | l | wasted |
|---|---|---|---|
| 7 | 5,040 | 13 | 62.5% |
| 8 | 40,320 | 16 | 62.5% |
| 9 | 362,880 | 19 | 44.5% |
| 10 | 3,628,800 | 22 | 15.6% |
| 11 | 39,916,800 | 26 | 68.1% |
| 12 | 479,001,600 | 29 | 12.1% |
| 13 | 6,227,020,800 | 33 | 37.9% |
| 14 | 87,178,291,200 | 37 | 57.7% |
| 15 | 1.30767E+12 | 41 | 68.2% |
| 16 | 2.09228E+13 | 45 | 68.2% |
| 17 | 3.55687E+14 | 49 | 58.3% |
| 18 | 6.40237E+15 | 53 | 40.7% |
| 19 | 1.21645E+17 | 57 | 18.5% |
| 20 | 2.4329E+18 | 62 | 89.6% |
| 21 | 5.10909E+19 | 66 | 44.4% |
| 22 | 1.124E+21 | 70 | 5.0% |
| 23 | 2.5852E+22 | 75 | 46.1% |
| 24 | 6.20448E+23 | 80 | 94.8% |
| 25 | 1.55112E+25 | 84 | 24.7% |
| 26 | 4.03291E+26 | 89 | 53.5% |
| 27 | 1.08889E+28 | 94 | 81.9% |
| 28 | 3.04888E+29 | 98 | 3.9% |
| 29 | 8.84176E+30 | 103 | 14.7% |
| 30 | 2.65253E+32 | 108 | 22.3% |

Table I contains the number of permutations ($n!$) and the length of the address $l$ in bits as a function of $n$. The last column shows the number of wasted addresses, because the full space in the addresses is almost never used. The length of the address and percent of wasted addresses as a function of $n$ is shown on Figure 3. For $n \in \{5, 10, 12, 22, 28, 29\}$ the number of wasted addresses is below 20%.

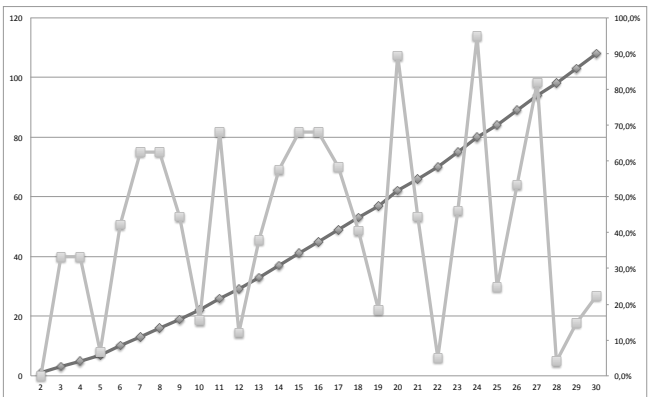

**Figure 3 Length of address and percent of wasted addresses as a function of $n$.**

Let us take a look at three examples. For two hashtags there are 2 bits for addressing with no wasted space and 2 permutations (Fig. 4). For three hashtags there are 3 bits (2 addresses wasted) and 6 permutations (Fig. 4). For four hashtags there are 5 bits for addressing with 8 wasted addresses and 24 permutations (Fig. 5).

| 0 0 0 0 0 | 1 2 3 4 | #alpha | #bravo | #charlie | #delta |
|---|---|---|---|---|---|
| 0 0 0 0 1 | 2 1 3 4 | #bravo | #alpha | #charlie | #delta |
| 0 0 0 1 0 | 1 3 2 4 | #alpha | #charlie | #bravo | #delta |
| 0 0 0 1 1 | 2 3 1 4 | #bravo | #charlie | #alpha | #delta |
| 0 0 1 0 0 | 3 1 2 4 | #charlie | #alpha | #bravo | #delta |
| 0 0 1 0 1 | 3 2 1 4 | #charlie | #bravo | #alpha | #delta |
| 0 0 1 1 0 | 1 4 2 3 | #alpha | #delta | #bravo | #charlie |
| 0 0 1 1 1 | 1 2 4 3 | #alpha | #bravo | #delta | #charlie |
| 0 1 0 0 0 | 2 4 1 3 | #bravo | #delta | #alpha | #charlie |
| 0 1 0 0 1 | 2 1 4 3 | #bravo | #alpha | #delta | #charlie |
| 0 1 0 1 0 | 1 4 3 2 | #alpha | #delta | #charlie | #bravo |
| 0 1 0 1 1 | 1 3 4 2 | #alpha | #charlie | #delta | #bravo |
| 0 1 1 0 0 | 2 4 3 1 | #bravo | #delta | #charlie | #alpha |
| 0 1 1 0 1 | 2 3 4 1 | #bravo | #charlie | #delta | #alpha |
| 0 1 1 1 0 | 3 4 1 2 | #charlie | #delta | #alpha | #bravo |
| 0 1 1 1 1 | 3 1 4 2 | #charlie | #alpha | #delta | #bravo |
| 1 0 0 0 0 | 3 4 2 1 | #charlie | #delta | #bravo | #alpha |
| 1 0 0 0 1 | 3 2 4 1 | #charlie | #bravo | #delta | #alpha |
| 1 0 0 1 0 | 4 1 2 3 | #delta | #alpha | #bravo | #charlie |
| 1 0 0 1 1 | 4 2 1 3 | #delta | #bravo | #alpha | #charlie |
| 1 0 1 0 0 | 4 1 3 2 | #delta | #alpha | #charlie | #bravo |
| 1 0 1 0 1 | 4 2 3 1 | #delta | #bravo | #charlie | #alpha |
| 1 0 1 1 0 | 4 3 1 2 | #delta | #charlie | #alpha | #bravo |
| 1 0 1 1 1 | 4 3 2 1 | #delta | #charlie | #bravo | #alpha |
| 1 1 0 0 0 | x x x x | | | | |
| 1 1 0 0 1 | x x x x | | | | |
| 1 1 0 1 0 | x x x x | | | | |
| 1 1 0 1 1 | x x x x | | | | |
| 1 1 1 0 0 | x x x x | | | | |
| 1 1 1 0 1 | x x x x | | | | |
| 1 1 1 1 0 | x x x x | | | | |
| 1 1 1 1 1 | x x x x | | | | |

**Figure 5 Example for $n = 4$.**

| 18 | #delta | #alpha | #bravo | **#charlie** |
|---|---|---|---|---|
| 20 | #delta | #alpha | #charlie | **#bravo** |
| 19 | #delta | #bravo | #alpha | **#charlie** |
| 21 | #delta | #bravo | #charlie | **#alpha** |
| 22 | #delta | #charlie | #alpha | **#bravo** |
| 23 | #delta | #charlie | #bravo | **#alpha** |
| 6 | #alpha | #delta | #bravo | **#charlie** |
| 10 | #alpha | #delta | #charlie | **#bravo** |
| 7 | #alpha | #bravo | #delta | **#charlie** |
| 0 | #alpha | #bravo | #charlie | **#delta** |
| 11 | #alpha | #charlie | #delta | **#bravo** |
| 2 | #alpha | #charlie | #bravo | **#delta** |
| 8 | #bravo | #delta | #alpha | **#charlie** |
| 12 | #bravo | #delta | #charlie | **#alpha** |
| 9 | #bravo | #alpha | #delta | **#charlie** |
| 1 | #bravo | #alpha | #charlie | **#delta** |
| 13 | #bravo | #charlie | #delta | **#alpha** |
| 3 | #bravo | #charlie | #alpha | **#delta** |
| 14 | #charlie | #delta | #alpha | **#bravo** |
| 16 | #charlie | #delta | #bravo | **#alpha** |
| 15 | #charlie | #alpha | #delta | **#bravo** |
| 4 | #charlie | #alpha | #bravo | **#delta** |
| 17 | #charlie | #bravo | #delta | **#alpha** |
| 5 | #charlie | #bravo | #alpha | **#delta** |

| for | **#delta** | go to | Facebook |
|---|---|---|---|
| | **#alpha** | | Google Plus |
| | **#bravo** | | Twitter |
| | **#charlie** | | Instagram |

**Figure 6 Example for $n = 4$ with addressing and pointers to social media networks.**

To start using Steghash we need to deal with four issues:
1. An algorithm for creating a dictionary – dependent only on $n$.

| 0 | 1 2 | #alpha | #bravo | |
|---|---|---|---|---|
| 1 | 2 1 | #bravo | #alpha | |

| 0 0 0 | 1 2 3 | #alpha | #bravo | #charlie |
|---|---|---|---|---|
| 0 0 1 | 2 1 3 | #bravo | #alpha | #charlie |
| 0 1 0 | 1 3 2 | #alpha | #charlie | #bravo |
| 0 1 1 | 2 3 1 | #bravo | #charlie | #alpha |
| 1 0 0 | 3 1 2 | #charlie | #alpha | #bravo |
| 1 0 1 | 3 2 1 | #charlie | #bravo | #alpha |
| 1 1 0 | x x x | | | |
| 1 1 1 | x x x | | | |

**Figure 4 Examples for $n \in \{2, 3\}$.**

2. A set of hashtags to create a dictionary.
3. The mapping of the addresses into a dictionary.
4. A secret transition generator to create the link between the addresses (a chain).

Any single sorting algorithm could be used – the choice of algorithm has no impact on the security if a secret transition generator (point 4) would be the pseudorandom. In 2 we need to balance the popularity of some hashtags and the freak to limit the search results. Typically one or two unpopular hashtags are enough to have a unique index. If all hashtags chosen for StegHash were popular we would need to look into each message from the search results to find the hidden content in the attached multimedia if present. A secret transition generator initiated with a secret password, as used in StegHash, produces addresses in a chain to go step by step. The first address is the start, and if we used all the space it would be similar to a circular linked list for the data structure. A secret transition generator is a function based on a pseudorandom code generator or a hash function.

| Step | From | | To | Addr |
|---|---|---|---|---|
| 1 | Facebook | → | Instagram | 18 |
| 2 | Instagram | → | Twitter | 20 |
| 3 | Twitter | → | Instagram | 19 |
| 4 | Instagram | → | Google Plus | 21 |
| 5 | Google Plus | → | Twitter | 22 |
| 6 | Twitter | → | Google Plus | 23 |
| 7 | Google Plus | → | Instagram | 6 |
| 8 | Instagram | → | Twitter | 10 |
| 9 | Twitter | → | Instagram | 7 |
| 10 | Instagram | → | Facebook | 0 |
| 11 | Facebook | → | Twitter | 11 |
| 12 | Twitter | → | Facebook | 2 |
| 13 | Facebook | → | Instagram | 8 |
| 14 | Instagram | → | Google Plus | 12 |
| 15 | Google Plus | → | Instagram | 9 |
| 16 | Instagram | → | Facebook | 1 |
| 17 | Facebook | → | Google Plus | 13 |
| 18 | Google Plus | → | Facebook | 3 |
| 19 | Facebook | → | Twitter | 14 |
| 20 | Twitter | → | Google Plus | 16 |
| 21 | Google Plus | → | Twitter | 15 |
| 22 | Twitter | → | Facebook | 4 |
| 23 | Facebook | → | Google Plus | 17 |
| 24 | Google Plus | → | Facebook | 5 |

**Figure 7 Transition graph for $n = 4$.**

As stated previously, a search engine designed for the OSNs or the interior search mechanism of the given OSN should be used to find the next messages. For some OSNs there are no effective search engines. We are able to take one hashtag or more as the pointer to the next OSN to increase the performance of the system. This has an impact on security, because the prediction of this type of subaddressing could be linked with a given OSN and could compromise the StegHash method. Figure 6 contains an example with four hashtags. The addressing scheme was taken from Figure 5 and then a SHA-512 [13] based function was used to produce a chain. The last hashtags in the index represent the placement of the next message (for X go to Y). Figure 7 shows a transition graph, which explains how a chain among the messages is built.

## IV. PROOF OF CONCEPT AND EVALUATION

We created a simple environment to prove that the concept of StegHash was proper (Fig. 8). The environment consisted of five components: four tools (StegDigger, StegHash Engine, StegPublishe, StegReader) and the database (DB).

As a carrier for the steganography we used pictures prepared with rules taken from the results of our previous effort [7]: we used pictures sanitized by the OSN, taken from the services directly with proper resolution and size. The StegDigger was responsible for collecting the content and storing the pictures in the DB. For the pictures stored in the DB we prepared several replicas with different steganographic algorithms and different sizes of embedded texts as hidden messages. StepDigger was just an overlay for a web browser working with publicly available profiles.

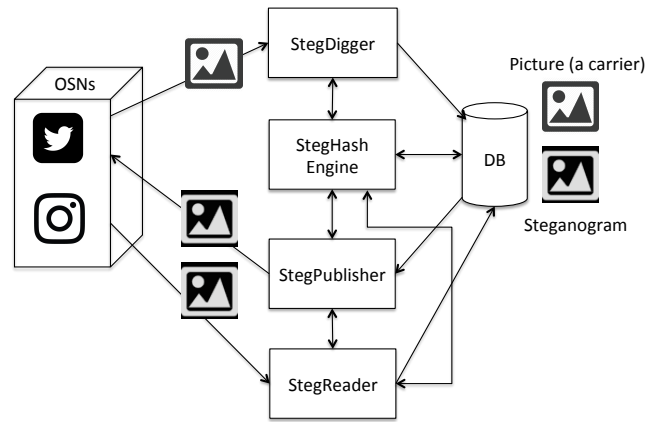

**Figure 8 Environment for evaluation.**

The StegHash engine was design to implement, as described in the previous chapter, the hashtag management and to connect the hashtags with hidden messages. StegPublisher was designed to work with two OSNs: Twitter and Instagram; similar to StegDigger the tool was just an overlay for a web browser. We noticed that Twitter has no limitation on searching for hashtags, but Instagram is limited to only one. Therefore, we decided to use a set of hashtags with rather uncommon words to give better performance when looking for a given message. We tested the process with several separate accounts to avoid being blocked by Twitter or Instagram due to massive traffic. Finally, StegReader was used for the evaluation of the message retrieval from the OSNs and it was physically integrated with StegPublisher.

We tested small sets of hashtags: $n \in \{3, 4, 5, 6\}$, so as not to interfere with the OSNs' performance and security policies. The experiment showed that the system worked, but as expected from the results presented in [7], we were not able to upload the all steganographic content with 100% accuracy and the average result was similar to that obtained in [7]. For short messages (up to 10 bytes of hidden data) our success rate was at 100%, but for longer messages (200-400 bytes) our success rate was 80%. We rebuilt our environment to improve the reliability: so after publishing every message, the system tried to recover the hidden part, and if it failed it was repeated by sending the message again with a new set of hashtags.

## V. Discussion

This paper is a report on work in progress rather than a publication of the final results, so there will now be a discussion in this section about some issues concerning the assumptions and the security of the proposed system.

In [14], a classification of steganography methods was presented with three levels of undetectability, named: "good", "bad", and "ugly". According to this categorization (which was formally proposed for network steganography, but that could be extended to all other methods with data in motion), StegHash seems to be a "good" method, as the observer is not able to detect the hidden communication anywhere in the network, even at the steganographic receiver of the hidden data.

In the experiment we did not use $n$ larger than 6, due to following the rules of (open) social coexistence, but it is of course possible.

The success rate of publishing the pictures with hidden messages depended on the algorithms used for the steganographic purposes on the client side, as well as on the algorithms for compressing the images on the server side. This is an area for future investigation, but from the functional perspective of the StegHash method it does not matter, as we could skip the failed messages.

The security of StegHash mainly depends on the proper management of the hashtags. From the OSNs perspective, the tracking of long sets of hashtags in messages with multimedia files with a rapid occurrence of such messages should be treated as anomaly behavior.

## VI. Conclusions and Future Work

In the initial experiment we proved the concept of the StegHash method, as a new approach for combining text steganography with other carriers, like pictures, movies, and songs, was correct. Please note that for $n$ hashtags, $m$ byte messages, and 100% accuracy, we have the receiving capacity of $n! \cdot m$ bytes for storage, i.e., for $n = 12$ and $m = 10$ bytes, this would be 4.46 TBytes. This is a promising use for StegHash, which can be like a FAT-equivalent (File Allocation Table). It seems that StegHash is a new hope for the time "before" post quantum cryptography by enabling the management of steganographic based storage.

In future work we will analyze other functions that have permutations for building relations among hashtags. In addition, we will use the OSNs' API (Application Programming Interface) rather than overlay methods for the software design. Finally, we are planning to use big data analytics to find the context in systems that are similar to StegHash.

Anyhow, it appears that StegHash opens OSNs to completely new kinds of threats, like grabbing a huge amount of storage, but simultaneously creates a new reason for the existence of social media.

## Acknowledgments

I would like to thank the following people for all their effort, motivation, and support: Miłosz Smolarczyk, Jędrzej Bieniasz, and Piotr Sapiecha.

This is an independent publication and has not been authorized, sponsored, or otherwise approved by Facebook, Google, Instagram, or Twitter.